\documentclass[prc,twocolumn,showpacs]{revtex4}
\usepackage{graphicx}
\usepackage{dcolumn}
\usepackage{bm}

\begin{document}

\title{E(5) and X(5) shape phase transitions within a Skyrme
 Hartree-Fock + BCS approach}

\author{R. Rodr\'{\i}guez-Guzm\'an and P. Sarriguren}

\email{sarriguren@iem.cfmac.csic.es}

\affiliation{Instituto de Estructura de la Materia, CSIC, Serrano
123, E-28006 Madrid, Spain}

\date{\today}

\begin{abstract} 
Self-consistent Skyrme Hartree-Fock plus BCS calculations are performed
to generate potential energy curves (PEC) in various chains of Pd, Xe,
Ba, Nd, Sm, Gd, and Dy isotopes. The evolution of shapes with the number
of nucleons is studied in a search for signatures of E(5) and X(5)
critical point symmetries. It is shown that the energy barriers in the
PECs are determined to a large extent by the treatment of the pairing 
correlations.
\end{abstract}


\maketitle

\section{Introduction}

The ground states of atomic nuclei are characterized by different
equilibrium configurations which correspond to different geometrical
shapes. The study of these equilibrium shapes, as well as the
transition regions between them, has been the subject of a large number
of theoretical and experimental studies (for a review, see, for example, 
Ref. \cite{review} and references therein).

Within the framework of algebraic models, the different nuclear
phase shapes are put in correspondence to dynamic symmetries of
some algebraic structure that links a specific mathematical
symmetry with a specific nuclear shape. Dynamic symmetries provide
a useful tool to describe properties of different physical systems
since they lead to exactly solvable problems and produce all results
for observables in explicit analytic form.

In nuclear physics, the algebraic structure of relevance, according
to the Interacting Boson Model (IBM) \cite{ibm}, is given by U(6).
There are three dynamical symmetries characterized by U(5), associated
to spherical symmetry, SU(3), associated with axially deformed 
symmetry, and SO(6), describing $\gamma$-unstable shapes. Experimental 
examples of all three types of symmetries have been recognized in many 
nuclei.

The phase shape transitions correspond to the breaking of these dynamic
symmetries and they occur as the number of nucleons change in the
nucleus. Understanding the behavior of systems undergoing a phase
transition is of especial relevance, since a complicated interplay of
competing degrees of freedom occurs at the critical points.

Iachello \cite{ia00,ia01} introduced the E(5) and X(5) critical point
symmetries within the framework of the collective Bohr Hamiltonian
\cite{bm} under some simplifying approximations. Critical point
symmetries provide parameter free (up to scale factors) predictions of
excitation spectra and electric quadrupole B(E2) strengths for nuclei
at the critical point of a phase/shape transition. The geometrical shape
of the ground state can be described \cite{bm} by three Euler angles and
by the quadrupole deformation parameters $\beta$ and $\gamma$. At the 
critical point of the phase transition the potential in the $\beta$ 
degree of freedom can be approximated by a simple square well potential, 
which is decoupled from the potential in the $\gamma$-variable.
In the case of the E(5) critical point symmetry, which corresponds to
the transition from spherical vibrational U(5) to deformed 
$\gamma$-unstable O(6), the potential is flat in the $\gamma$-direction. 
In the case of the symmetry X(5), related to the transition from U(5) 
to axially symmetric prolate SU(3), a harmonic oscillator potential is 
used in the $\gamma$-direction. 

Empirical evidence of these transitional symmetries at the critical
points were soon found in $^{134}$Ba \cite{cas00} for E(5) and in 
$^{152}$Sm  \cite{cas01} for X(5). Other nuclei have been also 
identified as good candidates for those symmetries. Such is the case
of $^{102}$Pd \cite{zamfir}, and $^{128-130}$Xe  \cite{clark}, which 
provide examples of E(5) symmetry. Other N=90 isotones like $^{150}$Nd  
\cite{krucken}, $^{154}$Gd \cite{tonev}, and $^{156}$Dy \cite{dewald} 
also provide examples of X(5) symmetry.

Algebraic models are very suitable for systematic studies because they 
provide powerful predictions with a very small number of parameters, 
but in order to deepen into the details, one has to perform microscopic
investigations of shape transitions and critical points which are,
to a large extent, still missing. In particular, it is 
interesting to examine whether the assumptions of relatively flat 
potentials in E(5) and X(5) are justified in different microscopic
models and the self-consistent mean-field approximation, based on 
parametrizations widely used all over the nuclear chart, appears as a 
very attractive initial tool for a link between algebraic models and 
microscopic theories. In this context, the relativistic mean-field
framework has been employed, in calculations of PECs as functions of 
the quadrupole deformation \cite{meng,sheng,fossion}. These studies 
have been performed for isotopic chains in which the occurrence of 
critical point symmetries has been predicted. Since flat PECs are one 
of the expected characteristics of critical point symmetries, 
constrained calculations in those isotopic chains should result in 
relatively flat PECs for nuclei with the critical symmetry.
It has been shown that particular isotopes exhibit relatively flat 
PECs over an extended range of the deformation parameter. Nevertheless, 
the behavior of the PECs, and particularly the potential barriers, are 
quite sensitive to the relativistic interaction used \cite{meng}, as 
well as to the pairing treatment. Therefore, the question arises whether 
a similar situation occurs when using nonrelativistic effective 
interactions to study  candidates for critical point symmetries. 
Such systematic nonrelativistic studies are still 
missing  and it is very interesting to compare the conclusions 
extracted from them with those obtained using both algebraic models 
and relativistic mean-field approximations.

In this work, we extend the calculations mentioned above to the 
case of  nonrelativistic self-consistent Skyrme Hartree-Fock + BCS 
mean-field calculations. The shape phase transitions corresponding 
to E(5) and X(5) symmetries are investigated systematically in 
various isotopic chains containing some of the suggested critical 
nuclei. In particular we study Pd, Xe, and Ba isotopic chains as 
candidates to E(5) symmetry, and  Nd, Sm, Gd, and Dy isotopic 
chains as examples of X(5).

The paper is organized as follows. In Sec. \ref{T-FRAMEWOK} we present 
a brief description of the theoretical formalism (Hartree-Fock + BCS)
used to obtain the main ingredient of the present study, i.e., the PECs
for the considered isotopic chains. For a more detailed account the 
reader is referred to the corresponding literature. Sec. \ref{Results} 
contains our results with a discussion on the sensitivity of the PECs 
to the effective nucleon-nucleon force and to the treatment of the 
pairing correlations. Sec. \ref{Conclusions} is devoted to the 
concluding remarks.

\section{Theoretical framework}
\label{T-FRAMEWOK}
The microscopic approach used in this work (i.e., HF+BCS) is based on
a self-consistent formalism built on a deformed Hartree-Fock (HF) 
mean-field, using Skyrme type energy density functionals. Pairing
correlations between like nucleons are included in the BCS approximation.
It is well known that the density-dependent HF+BCS approximation 
provides a very good description of ground-state properties for both 
spherical and deformed nuclei \cite{flocard} and it is at present one 
of the state-of-the-art  mean-field descriptions \cite{Bender-Review}.

There are two leading choices of methods to solve the deformed
HF+BCS equations. One option is the use of a coordinate space mesh. 
In this case one solves the HF+BCS equations for Skyrme type 
functionals via discretization of the individual wave functions on 
a 3-dimensional cartesian mesh \cite{ev8}. This corresponds to an 
expansion on a  specific basis of Lagrange polynomials associated 
with the selected mesh. The other common choice is to expand the 
single particle wave functions into an appropriate orthogonal basis 
(commonly the eigenfunctions of an axially symmetric harmonic 
oscillator potential). In the present study we perform calculations 
with both methods. We use the code {\bf ev8} \cite{ev8} in the first 
case and  follow the procedure based on the formalism developed in 
Ref. \cite{vautherin} in the second case.

In this work we consider the parametrization SLy4 \cite{sly4}
of the Skyrme force in the particle-hole channel, although we also
show results in some instances for the forces Sk3 \cite{sk3} and SG2
\cite{sg2}. They are examples of global effective interactions of
Skyrme-type that have been designed to fit ground state properties of
spherical nuclei and nuclear matter properties. While Sk3 is the most
simple one, involving in particular a linear dependence on the density, 
SLy4 is one of the most recent  parametrizations of Skyrme forces.

As we move away from closed shells, pairing correlations play an
important role \cite{rs} and should be taken into account. If one 
were dealing with a fundamental many-body Hamiltonian, one would 
proceed to apply Hartree-Fock-Bogoliubov formalism to it. However, 
dealing with Skyrme forces that have been simplified with the aim 
of reproducing average or bulk properties of the nucleus, one would 
have to include additional parameters in order to guarantee that 
sensible pairing matrix elements are obtained \cite{ev8,Bender-Review}.

\begin{figure}
\includegraphics[width=85mm]{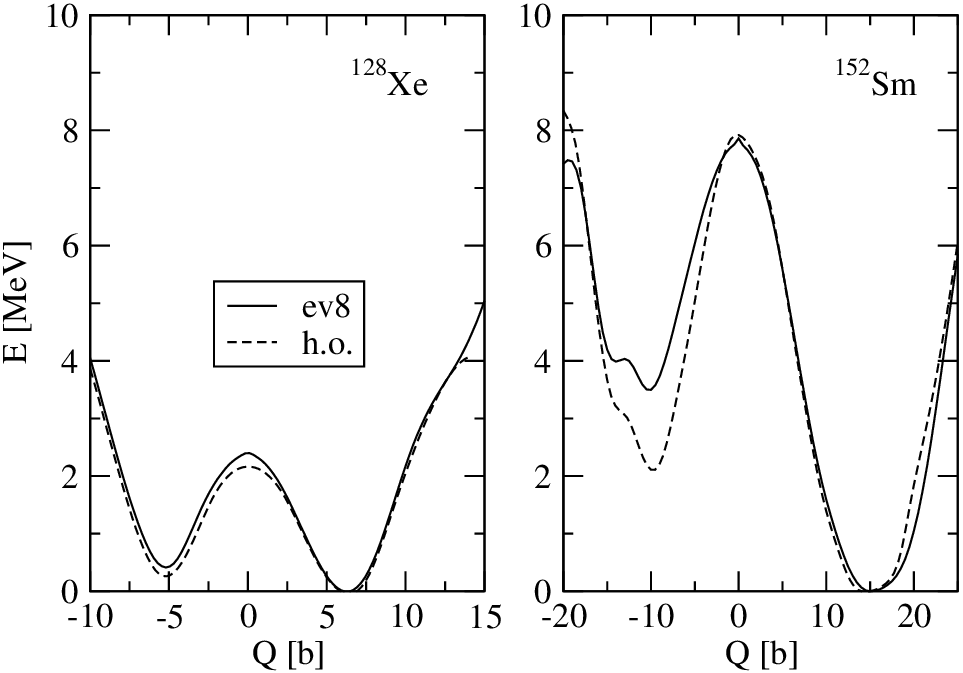}
\caption{Potential energy curves obtained with SLy4 and fixed pairing
gaps, using different methods for solving the HF+BCS equations.}
\label{fig1}
\end{figure}

\begin{figure}
\includegraphics[width=85mm]{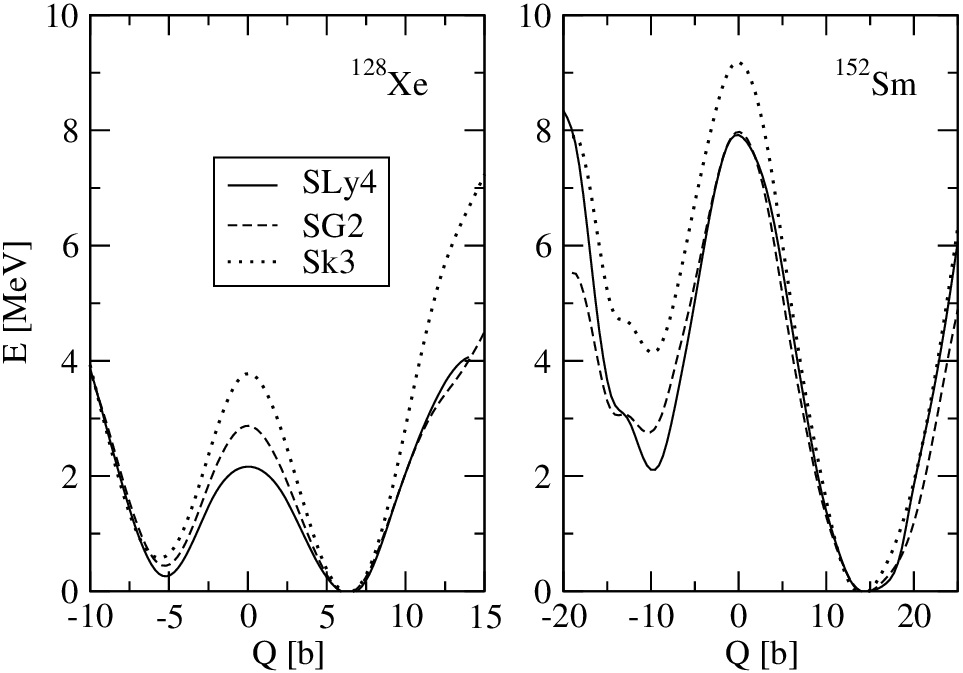}
\caption{Potential energy curves obtained with three different Skyrme
forces. For details see the main text.}
\label{fig2}
\end{figure}

\begin{figure}
\includegraphics[width=85mm]{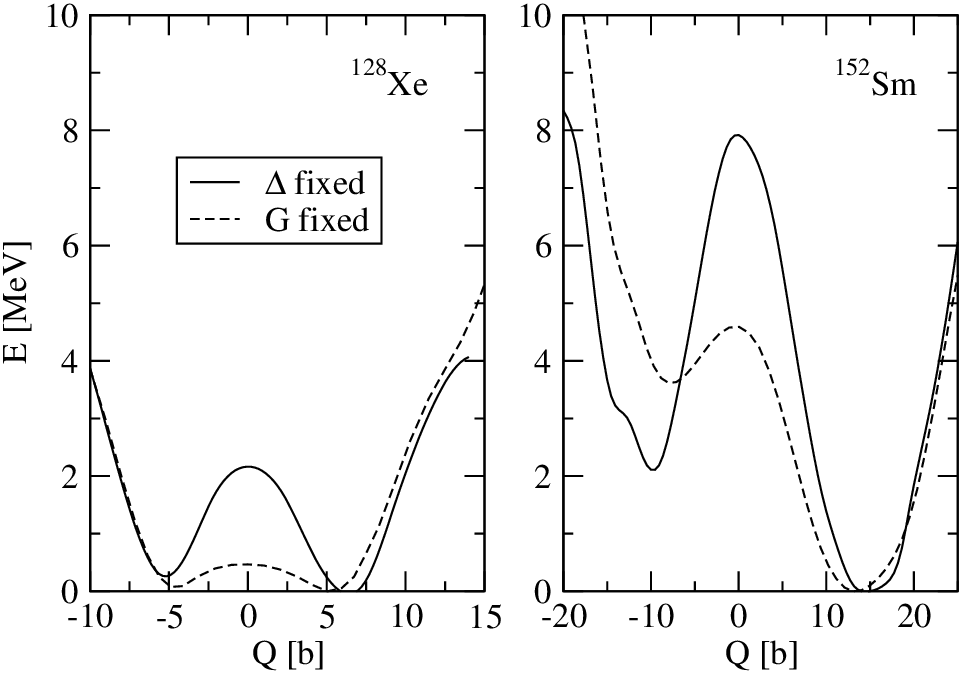}
\caption{Potential energy curves obtained with different treatments
of pairing. For details see the main text.}
\label{fig3}
\end{figure}

In this study pairing correlations are taken into account in the BCS 
approximation. Several options have been investigated. Our main 
option will be the use of a zero-range density-dependent pairing 
force \cite{terasaki}, 

\begin{equation}  
V({\bf r_1},{\bf r_2})=-g \left( 1-\hat{P}^{\sigma} \right)
\left( 1-\frac{\rho({\bf r_1} )}{\rho_c}\right)
\delta ( {\bf r_1}- {\bf r_2}  )\, ,
\label{dd-pairing}   
\end{equation}
where $\hat{P}^{\sigma}$ is the spin exchange operator, $\rho({\bf r})$
is the nuclear density, and $\rho_c=0.16$ fm$^{-3}$. The strength  $g$ 
of the pairing force [Eq.(\ref{dd-pairing})] is taken $g=1000$ MeV-fm$^3$  
for both neutrons and protons and a smooth cut-off of 5 MeV around the 
Fermi level has also been used. Let us mention that very recently the 
parametrization SLy4 has been successfully applied in combination with 
the pairing interaction [Eq.(\ref{dd-pairing})]  (with $g=1000$ MeV-fm$^3$)
in systematic studies of correlation energies from $^{16}$O to the 
superheavies \cite{other1} and in global studies of spectroscopic 
properties of the first $2^{+}$ states in even-even nuclei \cite{other2}.  
This is the main reason for selecting the combination SLy4 in the 
particle-hole channel and the  interaction [Eq. (\ref{dd-pairing})] 
(with $g=1000$ MeV-fm$^3$) in the pairing channel as the leading choice 
for the present study. Additionally, results with the strength  
$g=1250$ MeV-fm$^3$ will also be shown in some cases.

Another common practice to include pairing correlations is to introduce
a schematic seniority pairing force with a constant pairing strength 
$G$ \cite{rs} , we call this treatment constant-force approach,

\begin{equation}
V_{pair}=-G \sum_{m,m'>0} a_m^{\dagger} a_{\bar{m}}^{\dagger}
a_{\bar{m'}} a_{m'}\, .
\end{equation}
The strength of the pairing force for protons and neutrons $G_{p,n}$
is chosen in such a way that the experimental pairing gaps extracted
from binding energies in neighboring nuclei are reproduced. One can
further simplify the pairing treatment by parametrizing the pairing 
gaps $\Delta_{p,n}$ directly from experiment, we call this treatment 
constant-gap approach. The pairing strength and pairing gap are 
related through the gap equation \cite{rs}

\begin{equation}
\Delta = G \sum_{\nu >0} u_{\nu} v_{\nu} \, ,
\end{equation}
where $v_{\nu}$ are the occupation amplitudes.

The PECs shown in this study are computed microscopically by constrained
HF+BCS calculations \cite{ev8,rs,constraint}. These PECs  are obtained
by minimizing the corresponding energy functional under a quadratic 
constraint that holds the nuclear quadrupole moment fixed to a given 
value. In this work we show the energies as a function of the (axially 
symmetric) mass quadrupole moment $Q$, which is related with the 
quadrupole deformation parameter $\beta_2$ by the expression
\begin{equation}
\beta_2 = \sqrt{\frac{\pi}{5}}\frac{Q}{A<r^2>}\, ,
\end{equation}
in terms of the mean square radius of the mass distribution $<r^2>$.
Tables I and II contain the quadrupole parameters $\beta_2$
corresponding to the equilibrium configurations for the isotopes
considered in this work.

\begin{table}[h]
\begin{center}
\caption{Quadrupole deformation parameters $\beta_2$ for the
ground states in Pd, Xe, and Ba isotopes. SLy4 force and two
values of the pairing strength g [MeV fm$^{3}$] are used.} 
{\begin{tabular}{c|cccccccccc} 
\hline
\hline Pd & 96 & 98 & 100 & 102 & 104 & 106 & 108 & 110 & 112 & 114 \\
\hline
g=1000 & 0.10 & 0.12 & 0.15 & 0.l7 & 0.18 & 0.17 & 0.17 & 0.18 & 0.19 & -0.19 \\
g=1250 & 0.00 & 0.10 & 0.14 & 0.15 & 0.16 & 0.16 & 0.15 & 0.13 & 0.13 & 0.11 \\ \\
\hline 
Xe & 118 & 120 & 122 & 124 & 126 & 128 & 130 & 132 & 134 & 136 \\
\hline
g=1000 & 0.15 & 0.29 & 0.29 & 0.23 & 0.20 & 0.17 & 0.14 & 0.11 & 0.08 & 0.00 \\
g=1250 & 0.09 & 0.27 & 0.27 & 0.20 & 0.18 & 0.15 & 0.13 & 0.09 & -0.01 & 0.00 \\ \\
\hline 
Ba & 120 & 122 & 124 & 126 & 128 & 130 & 132 & 134 & 136 & 138 \\ 
\hline
g=1000 & 0.34 & 0.28 & 0.27 & 0.25 & 0.23 & 0.20 & 0.17 & 0.13 & 0.11 & 0.00 \\
g=1250 & 0.31 & 0.29 & 0.27 & 0.24 & 0.22 & 0.19 & 0.15 & 0.12 & 0.08 & 0.00 \\
\hline 
\hline
\end{tabular}}
\end{center}
\label{TABLE1}
\end{table}

\begin{table}[h]
\begin{center}
\caption{Quadrupole deformation parameters $\beta_2$
for the ground states in Nd, Sm, Gd, and Dy isotopes. SLy4 force 
and pairing strength g=1000 [MeV fm$^{3}$] are used.}
{\begin{tabular}{c|cccccccc} 
\hline
\hline Nd & 142 & 144 & 146 & 148 & 150 & 152 & 154 & 156 \\
\hline
& 0.00 & 0.10 & 0.14 & 0.20 & 0.28 & 0.29 & 0.29 & 0.30 \\ \\
\hline 
Sm & 144 & 146 & 148 & 150 & 152 & 154 & 156 & 158  \\
\hline
& 0.00 & 0.07 & 0.16 & 0.20 & 0.25 & 0.29 & 0.30 & 0.31  \\ \\
\hline 
Gd & 146 & 148 & 150 & 152 & 154 & 156 & & \\ 
\hline
& 0.00 & -0.05 & 0.17 & 0.22 & 0.25 & 0.28 & & \\ \\
\hline 
Dy & 148 & 150 & 152 & 154 & 156 & 158 & & \\ 
\hline
& 0.00 & -0.08 & -0.14 & 0.21 & 0.26 & 0.28 & & \\
\hline 
\hline
\end{tabular}}
\end{center}
\label{TABLE2}
\end{table}

\section{Discussion of results}
\label{Results}

It is known that PECs are sensitive to the effective nuclear force in
both relativistic \cite{meng} and nonrelativistic \cite{tajima,sarri}
approaches, as well as to pairing correlations \cite{tajima,sarri}. 
Thus, it is worth starting our discussion on PECs by studying this
sensitivity in our case. We perform this study on the example of
$^{128}$Xe and $^{152}$Sm, which are proposed candidates for E(5) 
and X(5) symmetries, respectively.

\begin{figure}
\includegraphics[width=65mm]{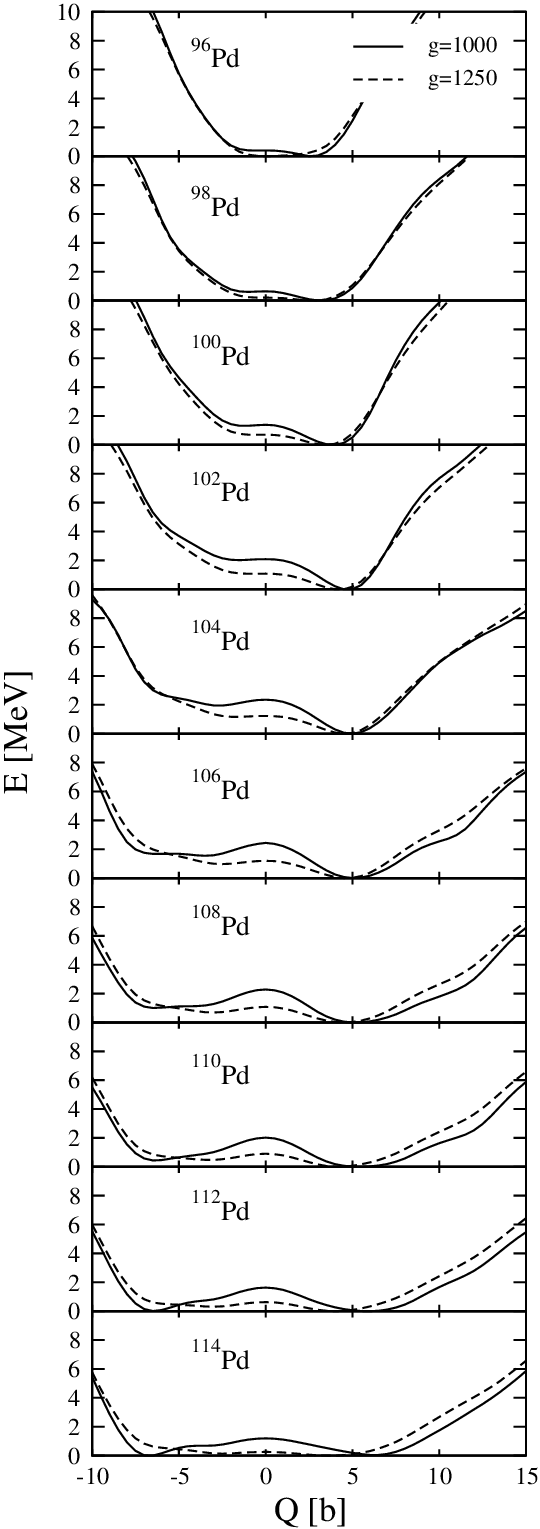}
\caption{Potential energy curves in Pd isotopes obtained from
constrained HF+BCS calculations with the force SLy4 and a zero-range
pairing force with two different strengths.}
\label{fig4}
\end{figure}

\begin{figure}
\includegraphics[width=65mm]{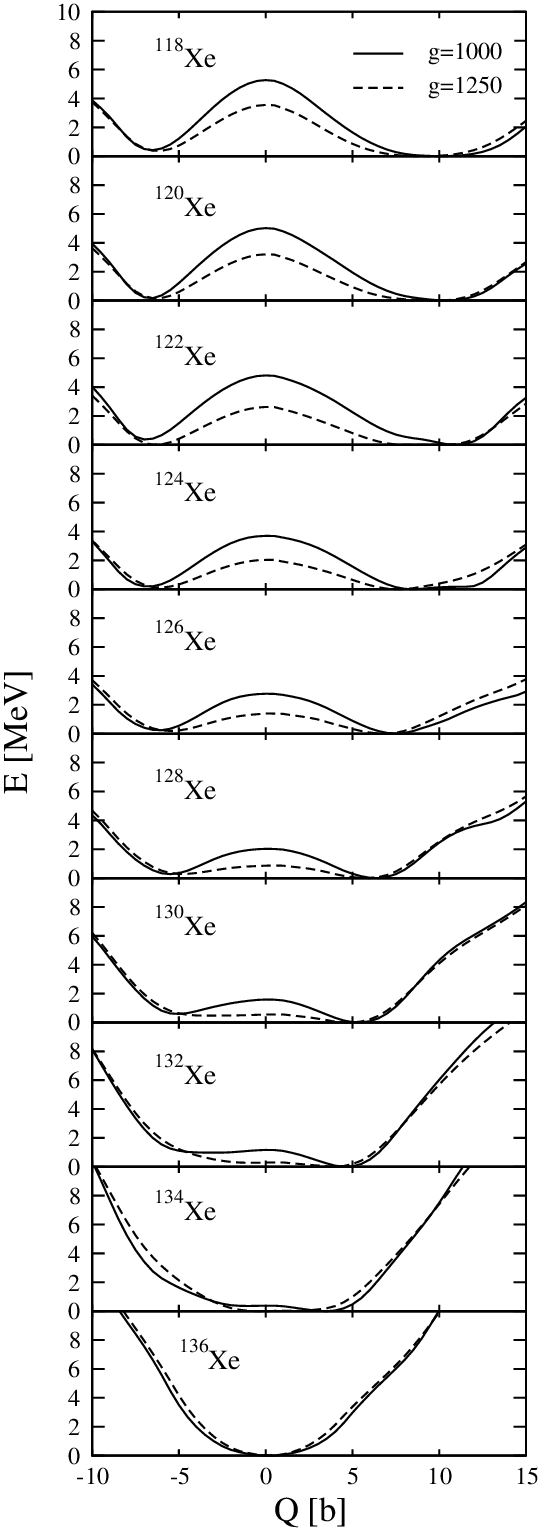}
\caption{
Same as in Fig.4 but for Xe isotopes.}
\label{fig5}
\end{figure}

\begin{figure}
\includegraphics[width=65mm]{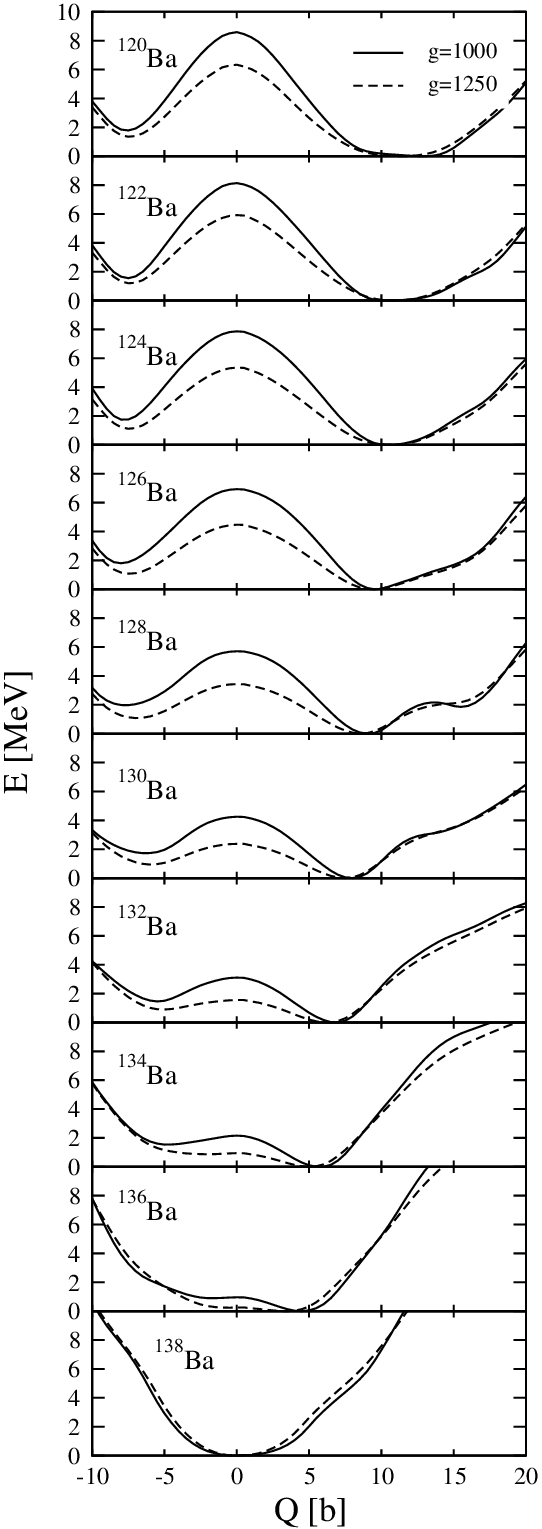}
\caption{
Same as in Fig.4 but for Ba isotopes.}
\label{fig6}
\end{figure}

In Fig. \ref{fig1} we study the effect of using different methods of 
solving the HF+BCS equations, coordinate lattice \cite{ev8} and deformed 
harmonic oscillator basis \cite{vautherin}. In these calculations we use 
the same SLy4 force in the particle-hole channel and the same constant-gap 
treatment of pairing. The results indicate that the PECs obtained are 
practically the same in these two cases and in general in all the
isotopes we have considered in this work.

In Fig. \ref{fig2} we consider the PECs obtained from three different 
Skyrme forces, namely SLy4, SG2, and Sk3, using the same pairing 
treatment. We can see that the location of the oblate and prolate minima 
appear at the same deformation no matter what the force is. However, the 
relative energies of the minima and the energy barriers between the 
minima can change by a few MeV, depending on the force.

In Fig. \ref{fig3} we show the effect of pairing using SLy4 force in all 
cases. We consider the constant-gap approach and the constant-force 
approach. Again, we observe that the minima appear at about the same 
deformations but the energy barriers can change considerably depending 
on the approach used.

From the analysis of Figs. \ref{fig1}, \ref{fig2} and \ref{fig3} we 
conclude that, at least in the mass region studied, the PECs are not 
sensitive to the method employed to solve the HF+BCS equations (3D 
cartesian lattice or deformed harmonic oscillator). We also conclude 
that the qualitative behavior of the energy profiles remains unchanged 
against changes in the Skyrme and pairing interactions in the sense 
that the deformations at which the minima occur are rather stable. 
Nevertheless, the relative energies of these minima, and particularly 
the energy barriers between them are  very sensitive to the details 
of the calculation, especially  to pairing. This is also the case in 
relativistic mean-field calculations. The sensitivity of the PECs to 
the force can be seen in Ref. \cite{meng}, where several forces were 
compared (NL1, NL3, NLSH, and TM1). The sensitivity to the pairing in 
relativistic calculations is also apparent if one compares the energy 
barriers in Sm isotopes obtained with the parametrization  NL3 in 
Ref. \cite{meng}, where a constant-gap approach was used, with those 
in Ref. \cite{fossion}, where a pairing based on the Brink-Boeker part 
of the Gogny force was used. One should remark that, at variance with
Ref. \cite{meng}, a self-consistent treatment of the pairing is used 
in Ref. \cite{fossion}. The main effect is a reduction of the energy 
barriers in the self-consistent treatment. In the case of nonrelativistic 
calculations with the Gogny interaction, since the same force is used in 
both mean-field and pairing channels, a fully self-consistent treatment 
is possible within a Hartree-Fock-Bogoliubov formalism. It will be very
interesting to explore whether this reduction of the barriers is also 
present. Work along this line is in progress.

To appreciate the effect on the energy barriers in more detail, we show
in Figs. \ref{fig4}, \ref{fig5}, and \ref{fig6} the results for PECs in  
$^{96-114}$Pd,  $^{118-136}$Xe, and $^{120-138}$Ba isotopic chains, 
respectively. The quadrupole deformations corresponding to the ground 
states of these isotopes can be found in Table I. These chains contain 
E(5) candidates found from systematic studies on available data on 
energy levels, E2, E1, and M1 strengths \cite{cas00,zamfir,clark,kir,arias}. 

The PECs shown in Figs. \ref{fig4}, \ref{fig5}, and \ref{fig6} have been 
computed within the mean-field scheme based on the Paris-Brussels code
{\bf ev8}, using the parametrization SLy4 and two different choices of 
the strength $g$ of the zero-range pairing force [Eq.(\ref{dd-pairing})], 
$g=1000$ MeV-fm$^3$ and $g=1250$ MeV-fm$^3$. From these figures one can 
see once more that oblate and prolate minima do not change significantly 
with the strength of pairing in each isotope. The only exceptions worth 
to mention are the cases  $^{96}$Pd, $^{114}$Pd, and  $^{134}$Xe, where 
we obtain the ground state at different deformations depending on the 
pairing strength $g$. This happens because of the flatness of the 
corresponding PECs, but in any case the energy difference between
the minima are always very small. On the other hand, the potential 
barriers are clearly lower when pairing is stronger. We can also observe 
the transition from the spherical isotopes $^{96}$Pd (N=50), $^{136}$Xe
(N=82), and $^{138}$Ba (N=82) to the $\gamma$-unstable isotopes 
$^{114}$Pd, $^{118}$Xe, and $^{120}$Ba. We identify the isotopes 
$^{108,110}$Pd, $^{128,130}$Xe, and $^{130,132}$Ba as transitional nuclei
with rather flat PECs (especially with g=1250). These results confirm 
the assumed square well potential in the $\beta$ degree of freedom
that leads to the critical point symmetry E(5). They are also
in good agreement with the results obtained in Ref. \cite{fossion} using 
the parametrization NL3 of the relativistic mean-field Lagrangian and 
a pairing force based on the Brink-Boeker part of the Gogny interaction.

\begin{figure}
\includegraphics[width=80mm]{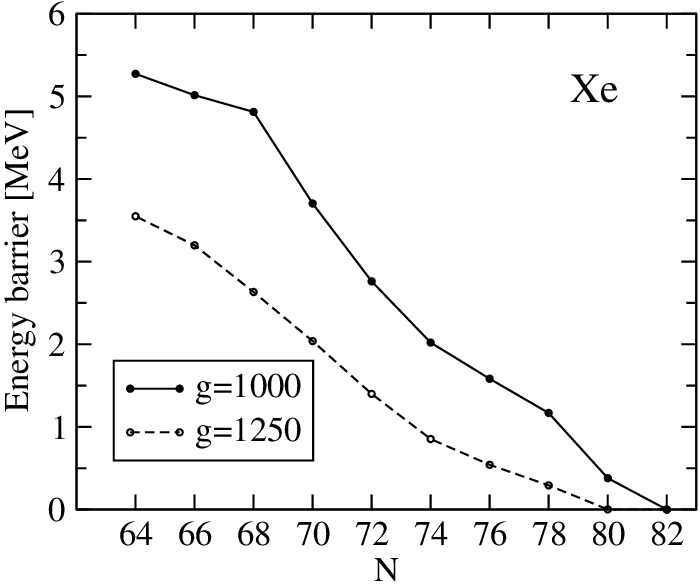}
\caption{Energy barriers in Xe isotopes corresponding to the cases 
shown in Fig. 5.}
\label{fig7}
\end{figure}

\begin{figure}
\centering
\includegraphics[width=75mm]{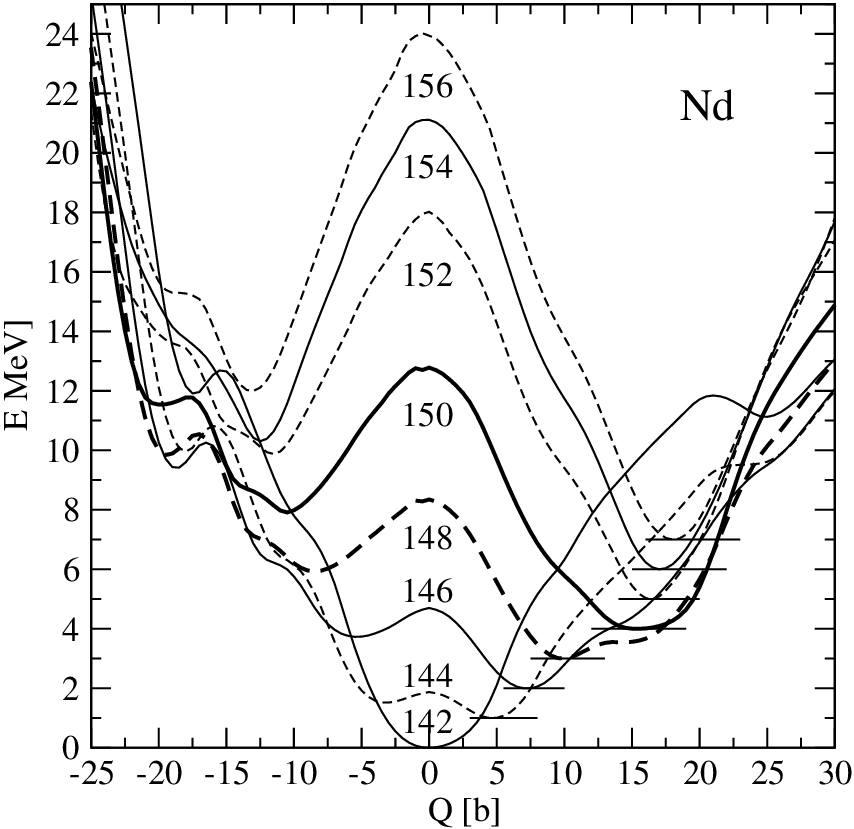}
\caption{Constrained HF+BCS calculations in $^{142-156}$Nd isotopes
with SLy4 and a zero-range pairing force with $g=1000$ MeV-fm$^3$. 
For a better comparison, the energies are shifted by 1 MeV for each 
isotope added, starting from A=142. Thick lines correspond to the 
isotopes that have been suggested to show a critical point symmetry.}
\label{fig8}
\end{figure}

Fig. \ref{fig7} shows the evolution of the energy barriers as a function 
of the number of neutrons for Xe isotopes for the two values of the
pairing strength $g=1000$ MeV-fm$^3$ and $g=1250$ MeV-fm$^3$ in 
[Eq.(\ref{dd-pairing})]. As it can be observed, the barriers obtained with 
the strength $g=1250$ MeV-fm$^3$ are systematically lower than the ones 
obtained with $g=1000$ MeV-fm$^3$. 

Now we turn the discussion to several rare-earth isotopic chains, Nd,
Sm, Gd, and Dy, where some of the nuclei (N=90 isotones) have been 
identified as exhibiting X(5) behavior, a transition between a spherical 
shape and a well deformed prolate shape. For example, the nucleus $^{152}$Sm  
\cite{cas01} was the first identified as exhibiting X(5) behavior and this 
is also the case for $^{150}$Nd \cite{krucken}. Other candidates have also 
been suggested with further work (see, for example, Refs. 
\cite{zamfir-other,clark-other,casten-other}).

Fig. \ref{fig8} shows the results for $^{142-156}$Nd isotopes (Z=60). The 
corresponding ground state deformation parameters for these nuclei are given 
in Table \ref{TABLE2}. We can observe a clear shape transition from spherical 
$^{142}$Nd (N=82) to clearly  prolate $^{152-156}$Nd. The isotopes $^{148}$Nd 
(N=88) and $^{150}$Nd (N=90) show a transitional behavior with a rather flat 
minimum on the prolate side and additional minima on the oblate sector at 
3 MeV and 4 MeV excitation energy, respectively. The energy barrier is about 
5 MeV in $^{148}$Nd and  8 MeV in $^{150}$Nd.
 
\begin{figure}
\centering
\includegraphics[width=75mm]{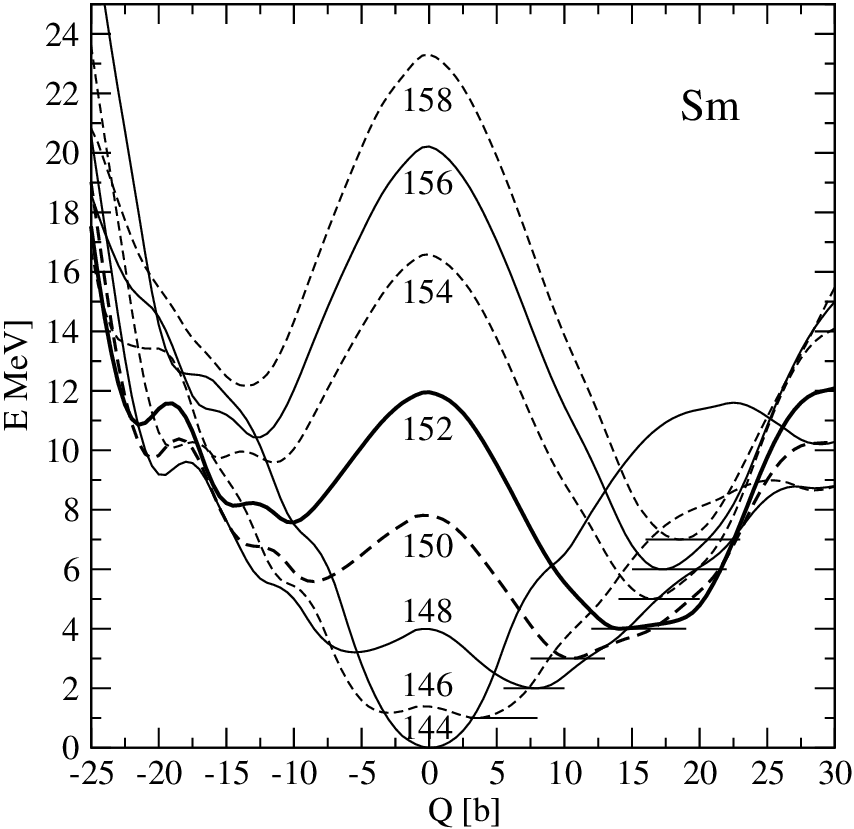}
\caption{Same as in Fig. \ref{fig8}, but for Sm isotopes.}
\label{fig9}
\end{figure}

\begin{figure}
\centering
\includegraphics[width=75mm]{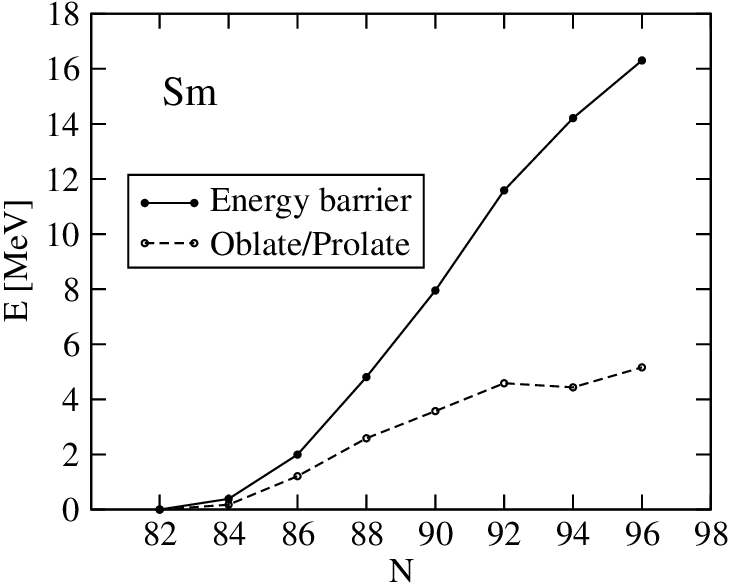}
\caption{Energy barriers and energy differences  between oblate and
prolate minima in the case of Sm isotopes.}
\label{fig10}
\end{figure}

\begin{figure}
\centering
\includegraphics[width=75mm]{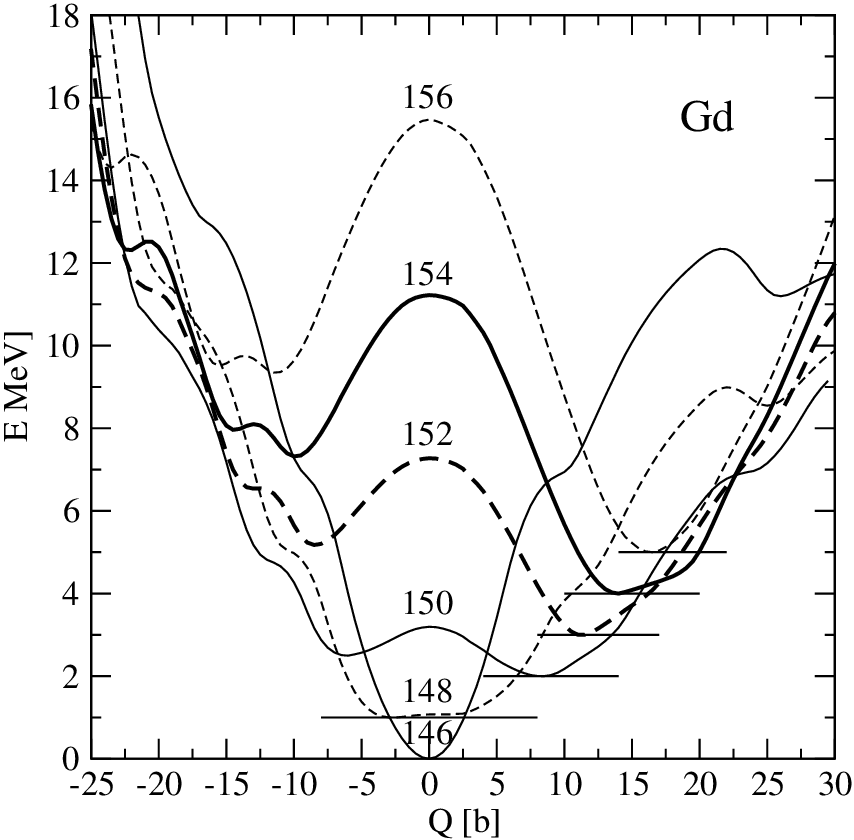}
\caption{Same as in Fig. \ref{fig8}, but for Gd isotopes.}
\label{fig11}
\end{figure}

\begin{figure}
\centering
\includegraphics[width=75mm]{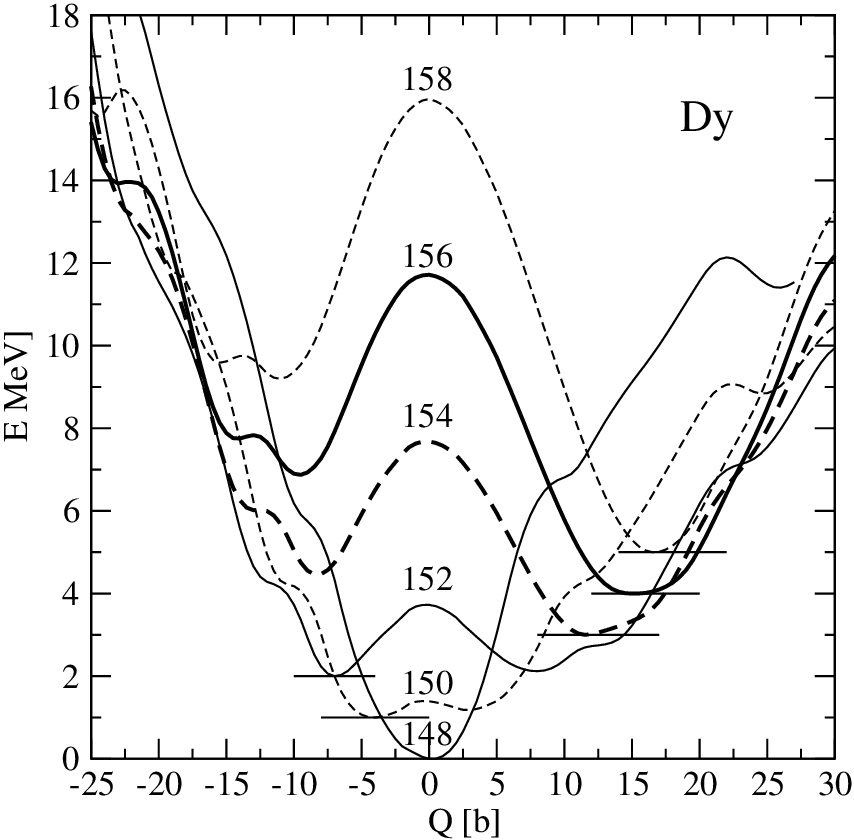}
\caption{Same as in Fig. \ref{fig8}, but for Dy isotopes.}
\label{fig12}
\end{figure}

Fig. \ref{fig9} contains the PECs for $^{144-158}$Sm isotopes (Z=62). The 
corresponding ground state deformation parameters  are given in Table 
\ref{TABLE2}. Similar arguments to those in Nd are valid. The transitional 
behavior between spherical, $^{144}$Sm, and well prolate deformed, 
$^{154-158}$Sm, appears again for the  N=88 and N=90 isotopes, $^{150}$Sm 
and $^{152}$Sm, respectively. The energy barriers and energy differences 
between oblate and prolate minima in Sm isotopes are shown in Fig. \ref{fig10}. 
They are qualitatively similar to the barriers obtained in the other 
rare-earth isotopic chains. Our PECs for $^{144-158}$Sm agree qualitatively 
with those of Refs. \cite{meng} and \cite{fossion}. They also agree well 
with the results obtained in Ref. \cite{Caprio-other} with  the 
Nilsson-Strutinsky + BCS calculations.

The same is true for Gd and Dy isotopes, whose PECs are shown in Figs. 
\ref{fig11} and  \ref{fig12}, respectively. The ground state deformation 
parameters for the considered Gd and Dy isotopes are given in 
Table \ref{TABLE2}. In this case, we find the transitional behavior for 
A=152,154 in Gd  isotopes and A=154,156 for Dy isotopes.

From the results described above we conclude that our calculations do 
not predict flat PECs for the N=90 isotopes, in agreement with 
previous studies (see, for example, Ref. \cite{fossion}). 

\section{Conclusions}
\label{Conclusions}

In this paper we have investigated the potential energy curves
in a series of isotopic chains containing transitional nuclei
that exhibit a critical point symmetry behavior.
A microscopic approach based on deformed HF+BCS calculations with
Skyrme forces has been used. 

We have studied the sensitivity of our results to the effective
nucleon-nucleon interaction and to the pairing force. From this 
analysis we conclude that the energy barriers in the PECs depend
strongly on the details of calculations, especially on the
pairing force. 
This is indeed relevant for this work because our purpose is to
study up to what extent the PECs exhibit a flat behavior at the
critical point symmetries.

We have found that the assumptions of flat potentials in the E(5)
critical point symmetry is supported by the present microscopic
calculations in  $^{108,110}$Pd, $^{128,130}$Xe, and $^{130,132}$Ba, 
that have been suggested as examples of E(5). In the case of X(5) 
we find that the rare-earth isotopes with N=88,90 show a 
transitional behavior that could be interpreted in terms of
X(5) symmetry. However,  we do not find a flat
behavior, in agreement with previous calculations
\cite{meng,sheng,fossion}. 

There remains a long list of tasks to  be undertaken  in the near future 
but the present study could be considered as a first step into a much 
more systematic exploration of the relation between algebraic models
and (nonrelativistic) microscopic models. In particular, it is very
important to understand how well the predictions of effective 
interactions with predictive power all over the nuclear chart compare
with those of the already mentioned algebraic models.

\begin{acknowledgments}
This work was partly supported by Ministerio de Educaci\'on y Ciencia
(Spain) under Contract No.~FIS2005--00640. One of us (R.R.) would like 
to thank Prof. F. Iachello (Yale University) for valuable suggestions 
and discussions.
\end{acknowledgments}

\end{document}